\documentclass[epsf]{aa}
\usepackage{graphicx}

\begin{document}
\def\fu{$f_1$}
\def\t{$\pm$}
\def\fd{$f_2$}
\def\ft{$f_3$}
\def\fq{$f_4$}
\def\fdu{$f_2 - 2f_1$}
\def\fp{$f_1 + f_2$}
\def\fm{$f_2 - f_1$}
\def\cd{d$^{-1}$}
\def\cds{d$^{-1}$\,}
\def\kms{km~s$^{-1}$}
\def\kmss{km~s$^{-1}$\,}
\def\I{\'\i}
\def\dsct{$\delta$~Sct}
\def\dscts{$\delta$~Sct\,}
\def\salp{\vskip 0.3truecm}
\title{Asteroseismology of HADS stars: V974 Oph, a radial pulsator
flavoured by nonradial components\thanks{Based on observations collected
at Europan Southern Observatory, La Silla, Chile}}
\author{E.~Poretti}
\institute {
INAF-Osservatorio Astronomico di Brera, Via Bianchi 46,
I-23807 Merate, Italy\\
 \email{poretti@merate.mi.astro.it}
}
\offprints{E. Poretti}
\date{Received 5 May 2003/ Accepted 7 July 2003}
\abstract{The analysis of a dense time--series on  
V974 Oph disclosed the rich pulsational content
(at least five independent terms) of this high--amplitude 
(0.60 mag in $B$--light) \dscts star. A mode with a
frequency very close to the main one (probably the fundamental
radial mode) has been detected: such
a doublet is not a common feature in stars of the same class. Also 
another  term can be considered a radial one, but the high ratio
(0.786) raises some problems that can be solved only by admitting
very low metallicity. It is quite evident that
some undetectable terms are again hidden in the noise, as
the least--squares fit leaves a rms residual much higher than
the observational noise. All that considered, nonradial modes
seem to play a key r\^ole in the light variability of V974 Oph.
Revealing an unsuspected asteroseismic interest, V974 Oph 
provides a link between low-- and high--amplitude \dscts stars.
}
\authorrunning{E. Poretti} 
\titlerunning{V974 Oph}
\maketitle

\section{Introduction} 
In one of the first reviews on \dscts and related stars, Breger (1979) proposed to
drop the distinction between high-- and small--amplitude pulsators located
on the lower part of the instability strip. This suggestion was applied  
by the GCVS, which simplified  the old nomenclature (RRs, Dwarf Cepheids,
AI Vel, \dsct,~...) only maintaining the subdivision between variables
belonging to the Pop.~I (\dscts stars, DSCT) and to the Pop.~II 
(SX Phe variables, SX PHE).  This simplification was widely accepted, as at
that time the amplitude was not considered a physical discriminant between
\dscts stars: the light curves of BN Cnc and AD CMi
shown by Breger (1979, his Fig.~1) are very similar in shape, but very different
in amplitude (0.014 and 0.294 mag, respectively), suggesting a different energy 
in the same physical process. 

The very strong observational effort made in the 80s and 90s demonstrated how
small amplitude DSCT stars show a large variety of nonradial modes and how their
light variability is complex (multiperiodicity, close doublets of frequencies,
amplitude variations,~...). They seem to be very different from high--amplitude
DSCT stars, which are classical radial pulsators, most monoperiodic. 
Therefore, the subclass of the High Amplitude Delta Scuti (HADS) stars
has been
introduced; their light curves have an asymmetrical shape which
can be quantified by Fourier decomposition. A review
of the HADS properties is given by McNamara (2000): it seems that HADS are
concentrated in the central part of the instability strip, in a well-defined
region. At first glance, the rough subdivision between radial, mostly monoperiodic 
(double--mode in some cases, but always radial modes) HADS stars and mostly nonradial,
multiperiodic low--amplitude stars seems valid. The acronym LADS has sometimes been
used to identify the latter variables, which are also considered as the
more promising asteroseismic targets. However, further investigations and
observational studies have modified this idyllic picture. Walraven et al. (1992)
found that AI Vel is a multiperiodic variable, probably in the fundamental and
first three overtone radial modes; they also suggest that a nonradial mode can
be present in the light curve of this HADS star. Later, Garrido \& Rodr\'{\i}guez
(1996) analyzed some time--series of HADS stars, finding that in the cases
of SX Phe and DY Peg other radial and nonradial modes could be excited. More
recently, Arentoft et al. (2001) found evidence both of amplitude  variation
and of excitation of nonradial modes in the light curve of V1162 Ori;
the same result has been obtained by Zhou (2002) on AN Lyn. Multimode SX Phe
variables have been found among the blue stragglers of NGC~3201 (Mazur et al. 2003).

Therefore, nowadays the phenomenology of HADS stars appears more similar to that
of  small--amplitude \dscts stars. The only remarkable difference is that
the nonradial modes in HADS stars have a much smaller amplitude (order of
magnitudes) than the radial modes and that period and amplitude variations can be 
considered as small perturbations of a mode always clearly visible in the light curve. 
In this paper we discuss the case of V974 Oph, which seems to make even closer
to each other the phenomenology of HADS and LADS stars. 

\section{Observations}
V974 Oph was monitored on 13 consecutive nights from April 11 to 23, 1989 as 
a target observed at the end of nights devoted to HD~101158 (Poretti 1991).
Measurements were performed at the ESO 50--cm telescope (La Silla, Chile --
now decommissioned)
by using the EMI~9789QB photomultiplier. HD 159822 has been used as comparison star
and the transformation into the standard system (by measurements of four stars
in the E7 region) supplied  $V$=10.08$\pm$0.01, $B$=10.23$\pm$0.01.
We obtained  1150 points on V974 Oph in $B$--light;  
the 528 measurements of
the check star HD 160108 yielded a standard deviation of 0.0041 mag in $B$--light.
\begin{figure}
\resizebox{\hsize}{!}{\includegraphics{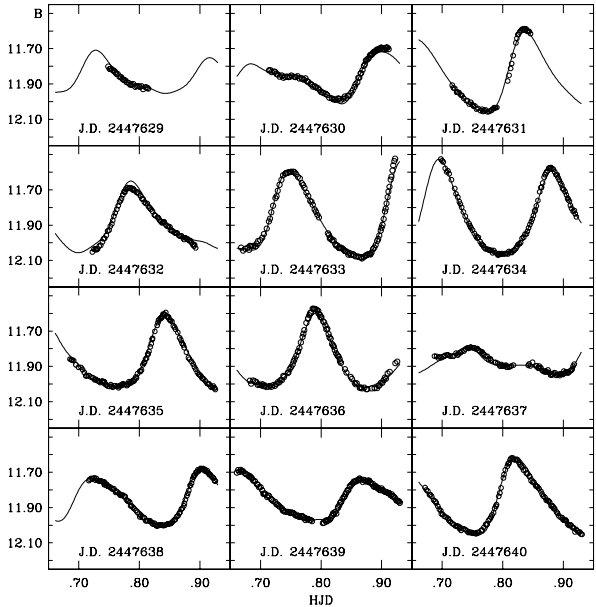}}
\caption[ ]{Light curves of V974 Oph in standard $B$ magnitudes. Solid line
represents the 12--terms fit.}
\label{curve}
\end{figure}

The observational history of V974 Oph is summarized by Poretti \& Antonello (1988).
The possibility that V974 Oph is similar to V1719 Cyg and V798 Cyg is 
ruled out when looking at the light curves obtained in this observing run
(Fig.~\ref{curve}).
\section{Frequency analysis}

\begin{table}
\caption{Parameters of the least--squares fit of the measurements on
V974 Oph. $T_0$=HJD~2447628.000}
\begin{tabular} {l c rrr r}
\hline
\multicolumn{1}{c}{Term} & & \multicolumn{1}{c}{Freq.} &\multicolumn{1}{c}{Ampl.} &
\multicolumn{1}{c}{Phase} &\multicolumn{1}{c}{$f_1/f_N$}\\
\multicolumn{1}{c}{} & & \multicolumn{1}{c}{[\cd]} &\multicolumn{1}{c}{[mag]} &
\multicolumn{1}{c}{[rad]} &\multicolumn{1}{c}{}\\
\hline
\noalign{\smallskip}
$f_1$        & &     5.2336 & 0.1587 & 2.55 & 1.000\\
$2f_1$       & &            & 0.0448 & 2.55 \\
$3f_1$       & &            & 0.0151 & 2.75 \\
$4f_1$       & &            & 0.0049 & 3.02 \\
$f_2$ & &     5.3498 & 0.0696 & 1.27 & 4.41 \\
$f_1+f_2$ & &           & 0.0149 & 4.36 \\
$2f_1+f_2$ & &           & 0.0062 & 4.82 \\
$f_3$        & &     6.6599 & 0.0427 & 5.96 & 0.786\\
$f_1+f_3$    & &            & 0.0119 & 6.02 \\
$f_3-f_1$    & &            & 0.0079 & 4.53 \\
$f_4$        & &     6.4762 & 0.0171 & 4.13 & 0.807\\
$f_5$        & &     7.1206 & 0.0104 & 0.61 & 0.734\\
\hline
\noalign{\smallskip}
\multicolumn{3}{l}{Mean $B$ magnitude}&\multicolumn{2}{c}{11.871$\pm$0.001}\\
\multicolumn{3}{l}{Residual rms}&\multicolumn{2}{c}{0.0141 mag} \\
\multicolumn{3}{l}{Measurements}&\multicolumn{2}{c}{1150} \\
\hline
\label{sol}
\end{tabular}
\end{table}

The least--squares power spectrum method (Vani\v{c}ek 1971)
allowed us to detect one by one the constituents of the light curve. After each
detection, we refined the frequency values by applying the MTRAP code (Carpino et
al. 1987) and they are introduced as known constituents (KCs) in
the new search. Such a procedure is  particularly suitable for a multimode, high--amplitude
 pulsation as it keeps the relationships between
the detected terms (i.e., 2$f_1$, $f_1+f_2$,~...) locked and it doesn't use
the data prewhitening, as amplitudes and phases of the KCs are recalculated
for each trial frequency.

\begin{figure}
\includegraphics[width=8.6truecm]{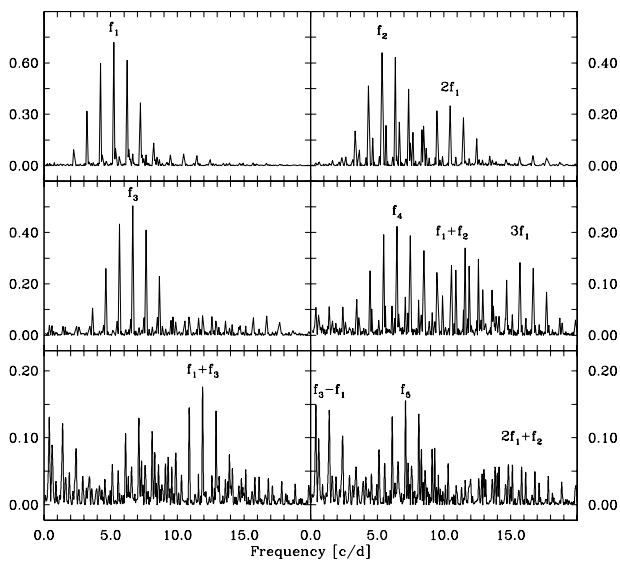}
\caption[ ]{Power spectra of the V974 Oph data.}
\label{high}
\end{figure}

The analysis of the time--series evidenced the large contribution of the $f_1$
term, i.e., 5.2336~\cds (Fig.~\ref{high}, top row, left panel). This periodicity
is characterized by a strong asymmetric light curve and the
harmonics 2$f_1$ (top row, right panel) and 3$f_1$ (middle row, right panel) are  
detected going further in the analysis. However, the
second contribution in amplitude comes from the $f_2$=5.3498~\cds term (top row,
right panel), i.e.,
a periodicity located close to $f_1$, just at the resolution limit
of our dataset. It 
should be noted that the alias at 6.3498~\cds is only slightly lower.  Since 
its amplitude is very large (0.07 mag), we can  verify if
the preference for the 5.3498~\cds term with respect to the 6.3498~\cds one is 
justified: indeed, when introducing the latter value as k.c., we recover a peak
at the former, confirming that our first choice is the right one. However, there
is another problem: being located at the resolution
limit, we cannot be sure that 5.3498~\cds is the exact value as 
its value can be modified by the interaction with $f_1$.

The other peaks clearly detectable in the power spectra are $f_3$=6.6599~\cds
and $f_4$=6.4762~\cds (middle row, left and right panels). Both these peaks
show a symmetric alias structure in the spectra: also for them, including an
alias as k.c. yields a residual signal at the central value, detectable later in
the frequency analysis.

In the fourth power spectrum we can also detect the structure related to the
11.58~\cd term; this term is the alias at +1~\cds of $f_1+f_2$ and its 
predominance over the expected value is not significant. Indeed, 
two steps beyond,
the $2f_1+f_2$ term is detected in the sixth power spectrum (bottom row, right panel).

The fifth power spectrum (bottom row, left panel) provides a good confirmation 
of the previous detections, as
the highest peak is at $f_1+f_3$=11.8935~\cds and the $f_3-f_1$=1.4236~\cds term 
is also visible, i.e., 
we detected both the coupling terms between $f_1$ and $f_3$. The last term
we can find in the time--series is $f_5$=7.1206~\cd, i.e., the highest peak in
the sixth spectrum.

Looking at Fig.~\ref{curve} we can expect that
a conspicuous set of frequencies is necessary to fit the light curves in a 
satisfactory way. 
We can propose five independent frequencies ($f_1$=5.2336~\cd, $f_2$=5.3498~\cd,
$f_3$=6.6599~\cd, $f_4$=6.4762~\cd, $f_5$=7.1206~\cd), 3 harmonics of $f_1$  and
 some
coupling terms between $f_1, f_2$ and $f_3$. Table~\ref{sol} summarizes the least--squares
solution (shown also in Fig.~\ref{curve}) of the $B$ magnitudes: a cosine
series has been used. 
Formal errors have been calculated following Montgomery \& O'Donoghue (1999):
they are around 4$\cdot$10$^{-4}$ mag on the amplitudes, 1--3$\cdot$10$^{-3}$\cds
on the frequencies of smaller amplitude, 10$^{-4}$\cds on
the frequencies of  larger amplitude, 10$^{-2}$~rad on the phases.
 
However, the least--squares solution is far from being considered as complete.
 The residual
rms is 0.015 mag, a value 3.6 higher than the observational error. However,
the residual rms doesn't decrease rapidly by adding new terms: indeed,
we get a rms residual of 0.0103 mag by introducing 18 KCs,
0.0086 mag by 24 k.c.'s, 0.0074 mag by 30 KCs. A huge
number of terms is necessary to reduce the rms residual to the noise level,
i.e., the 0.0041 mag obtained from the measurements of the
check star. 

\begin{figure}
\includegraphics[width=8.6truecm]{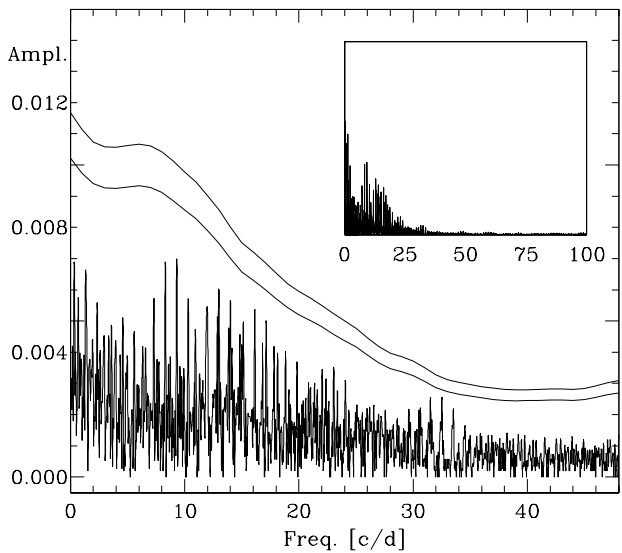}
\caption[ ]{Residual amplitude spectrum after having considered 12 known
constituents. Amplitudes are in mag. The upper line denotes S/N=4.0,
the lower line denotes S/N=3.5. The insert displays the power spectrum 
extended as far as  100~\cd.}
\label{resi}
\end{figure}

The power spectrum obtained considering the 12 terms of Tab.~\ref{sol}
as KCs is shown in Fig.~\ref{resi}.
Residual signal is present in different parts ($f<$2~\cd, $f\approx$~9~\cd, 
$f\approx$~13~\cd), but no peak reaches the acceptance levels ($S/N$=4.0 
for independent terms, $S/N$=3.5 for harmonics or combination terms;
Kuschnig et al. 1997).
As regards the coupling term $f_2-f_1$=0.1162~\cds   
we note that it is close to the resolution attainable with
our time baseline and therefore its non--detection is not surprising.  

\begin{figure}
\resizebox{\hsize}{!}{\includegraphics{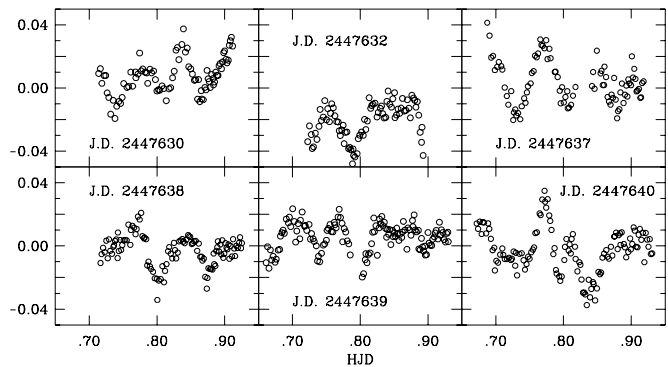}}
\caption[ ]{Residual after subtracting the 12-term solution from
the original measurements}
\label{oc}
\end{figure}

Figure~\ref{oc} shows the residuals between the measurements and the fitting
curve in few representative cases.  Most of the residual scatter mimics  high
frequency variations, but the signal seems confined in the low--frequency
region (see the insert in Fig.~\ref{resi}). As a matter of fact, the
power spectrum does not show any peak at all
for $f>25$~\cd. 
A possible explanation is the presence of other terms, not resolved
from the detected ones. In such a case, the least--squares solution 
can be locally unsatisfactory.

The residuals of JD 2447632 are systematically
shifted ($-$0.03 mag) respect with the zero level. This fact can suggest 
an instrumental effect,
but the measurements of the check star don't show any relevant peculiarity.
We also tried to fit the data by introducing individual night corrections,
but these arbitrary shifts didn't contribute to decrease rapidly the residuals. 

\section{Discussion}

The only tool we have to investigate
the pulsational content of V974 Oph is the ratio between the different
frequencies.  Recent observational studies allow us to extend the
same analysis to other HADS stars. 
Figure~\ref{ratio} shows the frequency content of V974
Oph compared to AI Vel (Walraven et al. 1992), V1162 Ori (Arentoft et
al. 2001), V567 Oph (Kiss et al. 2002), AN Lyn (Zhou 2002), SX Phe and
 DY Peg (Garrido \&
Rodr\I guez 1996), and MACHO stars (Alcock et al. 2000a). As regards
these latter stars, we considered only confirmed multimode pulsators
(i.e., stars listed in Tab.~4A in Alcock et al. 2002a,
but not those in Tab.~4B). In the figure we plot the observed $f_F/f_i$
ratio, assuming that the longest period is the fundamental radial
mode; the only exceptions are SX Phe and  MACHO 109.20634.24, where
the longest period is a small amplitude one, assumed to be a nonradial
mode. We also note that Garrido \& Rodr\I guez (1996) are rather
cautious on the reality of this period in the SX Phe data.
The canonical double--mode radial pulsators
have not been considered in Fig.~\ref{ratio}: they provide the well--known 0.77 ratio between
the fundamental (F) and the first overtone (1O) radial mode, shown as a vertical line.
On the other hand, double--mode pulsators with F/1O ratios slightly
different from 0.77 have been reported to show how these
discrepancies are currently observed (see Sect.~4.2).
The line corresponding to the F/2O ratio (i.e., 0.620; 
2O, second overtone radial mode) is also traced in Fig.~\ref{ratio}. We
have to note that the presence of small amplitude, nonradial
modes in the light curves of some stars has been established on
the basis of very different observational
conditions (time sampling, data homogeneity) and therefore they need
a more stringent observational confirmation.

The full amplitude of the V974 Oph light curve  is 0.60 mag in $B$--light, which scales
down to 0.43 mag in $V$ light. Among the stars considered here, only AI Vel and
MACHO 121.22427.551 (a canonical F/1O double--mode pulsator) 
have a higher amplitude (0.67 and 0.60, respectively);
SX Phe and V567 Oph have similar amplitudes, while the other stars are
around 0.20 mag. It should be noted that V974 Oph has the longest period, SX
Phe the shortest (0.055~d). Therefore, the sample is quite heterogeneous, also considering
that SX Phe is for sure a Pop.~II star, while others are more probably Pop.~I
stars.
\subsection{The $f_1-f_2$ doublet}
Close doublets of frequencies were recently found in different classes of high--amplitude
pulsators. They are common in RR Lyr stars, both fundamental (RRab, Moskalik \&
Poretti 2003) and first--overtone radial pulsators (RRc stars, Alcock et al.  2000b). 
There are also five well--defined examples (V1, V3, V4, V6 and V10) of close
doublets among the eleven SX Phe stars discovered in NGC~3201 (Mazur et al. 2003):
the distribution of the frequency ratios is well sharpened around 0.980,
with the only exception of the unclear case of V3.

In the case of V974 Oph we observe $f_1/f_2$=0.976, even if the $f_2$ term
can be affected by the limited frequency resolution (see Sect.~3).
A similar doublet (0.982)
has been found in the HADS star V1162~Ori (Fig.~\ref{ratio}). 
The frequency resolution of MACHO
data appears very suitable to detect close frequencies, but only 
MACHO 109.20634.24 shows a doublet composed of the
period of the $F$--mode and one slightly longer.
Close pair of frequencies are quite common among low--amplitude \dscts
stars (Breger \& Bischof 2002): however, in these stars  it is not possible to
assess the radial nature of one of the modes, owing to the large number of
detected modes. No close doublet has been detected in the light curve of AI Vel. 
The most plausible explanation for this is the excitation of a nonradial 
mode by resonance with the main period of oscillation.  

\subsection{The F/1O=0.77 ratio}
In the case of multimode HADS stars, it is quite obvious to  search for the ratio 0.77,
i.e., the theoretical F/1O ratio. At the beginning of the investigation on variability in
the lower part of the instability strip, such a search  was a misleading factor 
in the frequency analysis
of the low--amplitude \dscts stars, as their pulsational contents 
are actually more and more complicated. As a matter of fact, the 0.77
value was found between different terms solving the light curve of 44 Tau
(Poretti et al. 1992), demonstrating how nonradial modes can also originate it. 
On the other hand, such a ratio has been found  
in several high--amplitude radial pulsators (SX Phe, AE UMa, RV Ari, BP Peg and AI Vel). 

In the V974 Oph  data, the two frequencies $f_1$ and $f_3$ give $f_1/f_3$=0.786,
which appears a little higher value, but similar to those found  
in MACHO 121.22427.551 (0.783) and MACHO 114.20368.797 (0.790). 
V1162 Ori also supplies a still higher value (0.795).
The period ratios are very accurate: in the case of the shortest time baseline, i.e.
V974 Oph itself, the small errors on the frequencies propagate
on an  error bar of $\sim$1$\cdot$10$^{-4}$ on the ratio.
 The $f_1/f_3$ ratio raises
some theoretical problems. On one side, if  $f_1/f_3$ is recognized as the 
fingerprint of the F/1O ratio, V974 Oph
is an interesting laboratory to study the effects of the rotation and/or
the metallicity. Indeed, the high value requires  a very
low metallicity ($Z<$0.0005; see Fig.~3 in Petersen \& Christensen-Dalsgaard 1996)
and this seems to be unlikely. Fast rotation can
account for this value (P\'erez Hern\'andez et al. 1995).
However, some spectra taken with the Siding Spring 2.3--m telescope
show very narrow metallic lines, setting an upper limit of a few \kmss
(Kiss, private communication); hence, we can rule out fast rotation.
On the other hand, if $f_3$ is a nonradial mode, the resonance
effects ($f_1/f_2$ and $f_1/f_3$) are an intriguing theoretical point.

\subsection{The 0.800 ratio}
The 0.800 ratio is found in double mode Cepheids pulsating
in the 1O and 2O radial modes (Beltrame \& Poretti 2002). 
Such a value is also observed in some double--mode HADS stars (Musazzi et al. 1998)
and it is attributed to the same modes. In the case of V974 Oph, we  obtain 
$f_1/f_4$=0.807.  However, if we accept the previous identification
(i.e.,$f_1/f_3=$F/1O), we must consider $f_4$ as  a nonradial
mode, as it has an intermediate value between the frequencies of the $F$ 
 and 1O  modes. As a couple of values significantly different from 0.800 
are found in  MACHO 109.20634.24, the nonradial nature of some of these
modes seems assured.

\begin{figure}
\resizebox{\hsize}{!}{\includegraphics{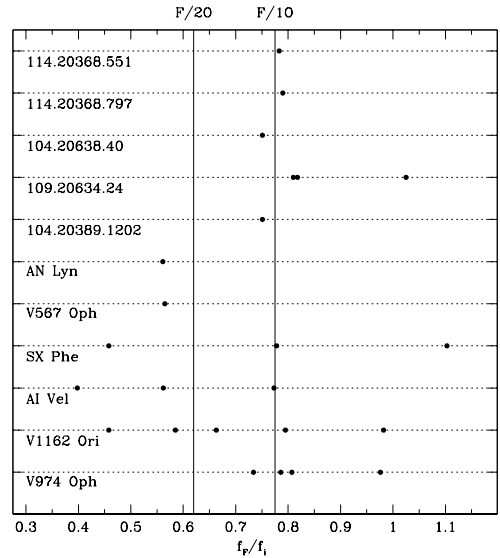}}
\caption[ ]{Observed (filled dots)  frequency ratios
 among HADS stars. Canonical F/1O pulsators are not shown.
Numbers indicate stars in the MACHO catalogue. Vertical
lines indicate the theoretical ratio between fundamental (F) and overtone
(1O, first; 2O, second) radial modes.}
\label{ratio}
\end{figure}
\section {Conclusions}
V974 Oph can be considered a unique high amplitude pulsating star in the 
current observational scenario:
\begin{enumerate}
\item The $f_1/f_2$ doublet makes V974 Oph similar to 
low--amplitude \dscts and SX Phe stars from short--period side and 
to RR Lyr variables from the long--period side. No other star with a
period around 0.20~d shows such a feature. 
\item The $f_1/f_3$ ratio seems a bit higher than expected for F and 1O 
radial modes. A detailed 
spectroscopic analysis could clarify this point, via an evaluation of the
influence of the metallicity or the detection of line profile
variations suggesting nonradial modes; 
\item Only  a large number of terms could reduce the observed scatter.
This finding looks very attractive for asteroseismic purposes, as it suggests
the excitation of nonradial modes, whose presence was not so evident in other
HADS stars. 
\end {enumerate}
In a certain sense, V974 Oph provides a link between low--
and high--amplitude variables. 

The identification of $f_3$ as a nonradial mode is also 
suggested by the presence of numerous cases of stars with frequency ratios
greater (see Sect.~4.2) and smaller 
(MACHO 104.20389.1202 and MACHO 104.20638.40, Fig.~\ref{ratio}) than the canonical
0.77--0.78 value. As in the case of the LADS, we should not consider such
a value uniquely as the signature of radial modes. If we also consider
the cases with ratio around 0.800, we can see how a large range of frequency ratios
is actually observed, as expected from the excitation of a large variety of nonradial modes. 
Figure~\ref{ratio} also shows that there are some HADS stars displaying a frequency ratio close
to 0.56. The most robust case is provided by the long--term photometry of AI Vel:
in their discussion Walraven et al. (1992) ruled out that such a ratio can be explained
by two radial modes. 

In the hypothesis of nonradial modes, it is quite interesting
to verify if FG Vir, i.e., a case study in the \dscts scenario, displays
the same ratios.
The fundamental radial mode (12.154~\cd) is well--identified and a  high number of nonradial
mode is observed (Breger et al. 1999, Mantegazza \& Poretti 2002). Indeed,
a doublet is observed, even if with a  slightly lower value
(12.154/12.716=0.956) than that  
observed in V974 Oph and other HADS stars. No frequency couple matches the 
0.77--0.78 ratio; there are three frequencies (21.052, 21.232, 21.551~\cd)
supplying a  ratio  in the 0.577--0.564 range, but none close to the 0.561 value. 
Different evolutionary stages (FG Vir is close to the Terminal Age Main Sequence, HADS
stars should be more evolved) can account for these differences. Anyway, it is
evident that asteroseismology is not confined in low amplitude pulsators and HADS stars
can provide useful, complementary observational facts to model the pulsation. On the
basis of the new results, the characterisation of the light variability  
of HADS stars needs to be done in a deeper way in the future. 

\begin{acknowledgements}
The research has made use of the {\sc simbad} database, operated at CDS, Strasbourg,
France. J.~Vialle improved the English form of the manuscript.
The author wishes to thank J.O.~Petersen for fruitful discussions at an
early stage of this work, for further  useful comments on a first draft
of the manuscript and on some checks about frequency analysis and modelling.
The author also wishes to thank Rafa Garrido for useful comments and
Laszlo Kiss for the spectroscopic determination of the $v\sin i$ value.
\end{acknowledgements}

\end{document}